\documentclass[11pt,a4paper]{article}
\usepackage{jstyle}

\newcommand{\bi}{\begin{itemize}}
\newcommand{\ei}{\end{itemize}}
\newcommand{\bea}{\begin{align}}
\newcommand{\eea}{\end{align}}
\newcommand{\be}{\begin{equation}}
\newcommand{\ee}{\end{equation}}

\newcommand{\gad}{{\dot{\alpha}}}
\newcommand{\gbd}{{\dot{\beta}}}

\newcommand{\ga}{\alpha}
\newcommand{\gb}{\beta}

\newcommand{\pl}{{\partial}}

\newcommand{\bry}{{{\bar{y}}}}

\newcommand{\brxi}{{{\bar{\xi}}}}

\newcommand{\breta}{{{\bar{\eta}}}}
\newcommand{\brzeta}{{{\bar{\zeta}}}}

\newcommand{\adD}{{D}} 
\newcommand{\tadD}{{\widetilde{D}}} 

\makeatletter
\renewcommand*\env@matrix[1][\arraystretch]{%
  \edef\arraystretch{#1}%
  \hskip -\arraycolsep
  \let\@ifnextchar\new@ifnextchar
  \array{*\c@MaxMatrixCols c}}
\makeatother

\author{\quad Massimo TARONNA$^{a}$ \footnote{Postdoctoral Researcher of the Fund for Scientific Research-FNRS Belgium.}}

\affiliation{${}^{a}$Physique Th\'eorique et Math\'ematique\\
\hspace*{0.01cm} Universit\'e Libre de Bruxelles
and International Solvay Institutes\\
\hspace*{0.01cm} ULB-Campus Plaine CP231, 1050 Brussels, Belgium}

\emailAdd{massimo.taronna@ulb.ac.be}

\title{\centering
\LARGE{A note on field redefinitions and higher-spin equations}}

\abstract{In this note we provide some details on the quasi-local field redefinitions which map interactions extracted from Vasiliev's equations to those obtained via holographic reconstruction. Without loss of generality, we focus on the source to the Fronsdal equations induced by current interactions quadratic in the higher-spin linearised curvatures.}

\begin{document}

\maketitle

\section{Introduction and Summary of Results}

The problem of admissible functional-classes has been of recent interest in the context of higher-spin (HS) theories \cite{Vasiliev:1990en}. In particular, in \cite{Kessel:2015kna,Boulanger:2015ova} the quadratic interaction term sourcing Fronsdal's equations was extracted from Vasiliev's equations obtaining an expression of the schematic form:\footnote{Notice that in the following we use the schematic notation $\Box^l\sim \dots\nabla_{\mu(l)}\phi\ldots\nabla^{\mu(l)}\phi$. We give precise formulas for the above contractions in the spinor language in the following section in eq.~\eqref{dic}. In this section all formulas are schematic and provide some intuition on their generic structure.}
\begin{equation}\label{back}
\Box\phi_{\mu(s)}+\ldots=\sum_{l=0}^\infty\frac{j_l}{l!l!}\,\Box^l\left(\nabla\ldots\nabla\phi\ \nabla\ldots\nabla\phi\right)_{\mu(s)}\,,
\end{equation}
which is sometime in the literature referred to as pseudo-local or quasi-local, meaning that it is a formal series in derivatives.\footnote{This terminology originates from the fact that formal series allow truncations to finitely many terms which are always local.}
The extracted Fronsdal current have coefficients whose asymptotic behaviour is given by $j_l\sim \frac1{l^3}$ for $l\to \infty$ for any choice of the spin $s$. The asymptotic behaviour of the coefficients raised the important question whether the backreaction extracted is or not strongly coupled\footnote{Preliminary questions of this type were raised in \cite{Boulanger:2008tg}.}. Furthermore, a key question whose study was undertaken in \cite{Skvortsov:2015lja,Taronna:2016ats} was whether it is possible to extract the coefficients of the canonical Metsaev vertices with finitely many derivatives \cite{Metsaev:2005ar,Joung:2011ww} from the above tails. Indeed most of the coefficients at the cubic order are unphysical, since they can be removed by local field redefinitions. In this respect, the full list of Metsaev-like couplings was indeed extracted holographically in \cite{Sleight:2016dba} and amounts to a finite number of coefficients for any triple of spins, to be contrasted to the above infinite set (see also \cite{Taronna:2010qq,Sagnotti:2010at} for the analogous string theory computation and corresponding cubic couplings).\footnote{It is important to stress that in a fully non-linear HS theory it is expected that the appropriate field frame which makes HS geometry manifest will entail \emph{all} of the above coefficients. The situation should be similar to the Einstein-Hilbert cubic couplings which are dressed by improvement terms that can be removed by a field redefinition at the cubic order. This is a further key motivation to understand these higher-derivative tails.}

Remarkably, the pseudo-local nature of the above currents implies that the only way to relate them to their Metsaev-like counterparts is via a pseudo-local field redefinition of the same schematic form:
\begin{align}\label{pseudored}
\phi_{\mu(s)}\rightarrow\phi^\prime_{\mu(s)}=\phi_{\mu(s)}+\sum_{l=0}^\infty \frac{a_l}{l!l!}\,\Box^l\left(\nabla\ldots\nabla\phi\ \nabla\ldots\nabla\phi\right)_{\mu(s)}\,,
\end{align}
involving a sum over infinitely many terms unbounded in derivatives (i.e. pseudo-local).
This result has motivated a renewed interest in the analysis of the admissible functional classes in HS theories. Indeed, an arbitrary pseudo-local redefinition defined in \eqref{pseudored} is sufficient to remove all pseudo-local current interactions \cite{Prokushkin:1999xq,Kessel:2015kna} and some further condition on the coefficients $a_l$ should be imposed on top of quasi-locality. A proposal\footnote{The proposal of \cite{Taronna:2016ats} is based on jet space and on the convergence of the infinite derivative expansion. It turns out that this proposal ensures the invariance of the Witten diagrams under the corresponding admissible field redefinitions.} based on the invariance of the holographic Witten-diagrams was put forward in \cite{Skvortsov:2015lja,Taronna:2016ats}, while in \cite{Vasiliev:2016xui} (see \cite{Vasiliev:2015wma} for further details) it was proposed to study classes of functions in $z$ and $y$ oscillators which are closed under star product multiplication. The aim of this note is to elaborate on these results from various perspectives, and to present the explicit form of the field redefinitions mapping the pseudo-local back-reaction \eqref{back} to its canonical (local) form.



In the following we list/summarise some relevant points of this analysis, together with the main results of this note, leaving the details of the derivation to the following sections:

\begin{enumerate}
\item Defining the canonical $s$-derivative current made out of two scalars as:
\begin{equation}\label{canJ}
J^{\text{can}}_{\mu(s)}=i^s\,\phi\,\overset{\leftrightarrow}{\nabla}_{\mu(s)}\phi\,,
\end{equation}
the redefinition which allows to bring the pseudo-local backreaction \eqref{back} to its canonical form:
\begin{equation}\label{caneq1}
\underbrace{\Box\phi_{\mu(s)}+\ldots}_{\mathcal{F}_{\mu(s)}}=\alpha_s \,J^{\text{can}}_{\mu(s)}\,,
\end{equation}
has the structure \eqref{pseudored} with the following choice of coefficients:
\begin{equation}\label{genred}
a_{l-1}=\frac{2l+s+2}{2}\left(\frac{l!}{(l+s+1)!}\right)^2[p_s(l)+\#\, \alpha_s]\,.
\end{equation}
Above $p_s(l)$ a polynomial of degree $2(s-1)$ in $l$ while we have left the coefficients $\alpha_s$ arbitrary. As detailed in the following sections the above discussion generalises to a current involving HS linearised curvatures of any spin. In this way it is transparent to compare redefinitions that give different answers for the overall coefficient of the canonical current in \eqref{caneq1}. Moreover, only one of the redefinitions considered above gives the coupling constant which matches the one derived in \cite{Bekaert:2015tva,Sleight:2016dba}. Keeping track of field normalisations (see Appendix~\ref{normaliz}), in the type A theory the choice expected from holography is:
\begin{equation}\label{recon}
\alpha_s=2^{1-s}\,g^2\,N_0^2\,,
\end{equation}
which has a simple spin-dependence up to a spin-independent factor proportional to the normalisation of the scalar field kinetic term and to the HS coupling constant $g$.

The leading asymptotic behaviour for $l\to\infty$ of the coefficients in the field redefinition \eqref{genred} is spin-independent and equal to $\frac{1}{l^3}$. Therefore, all the redefinitions considered above belong to the \emph{same} functional space as the current in eq.~\eqref{back} itself.

\item The value \eqref{recon} for the coupling constant can be fixed using Noether procedure to the quartic-order (see e.g. \cite{Kessel:2015kna} for the 3d computation using admissibility condition) and was reconstructed from Holography in \cite{Bekaert:2015tva,Sleight:2016dba}. So far, however, it was not possible to fix all cubic couplings using only the Noether procedure. That this should be possible in principle is suggested by the result of Metsaev in the light-cone gauge \cite{Metsaev:1991nb,Metsaev:1991mt}.\footnote{In covariant language cubic couplings, including highest derivative ones, should be fixed by global part of the HS symmetry from the equation $\delta^{(1)}S^{(3)}\approx0$.} Furthermore, the implications of the field redefinition mapping the theory to its canonical form at cubic order should be analysed at the quartic order. Such non-local redefinitions would generate a non-local quartic coupling. To appreciate the issue it might be worth noting that the above redefinitions can generate quartic couplings which differ from each other by single-trace blocks in the corresponding conformal block expansion (see e.g. \cite{Bekaert:2014cea,Bekaert:2015tva,Taronna:2016ats,Bekaert:2016ezc,Sleight:2016hyl}). It is also conceivable that, at the quartic order another non-local redefinition will be needed to compensate the cubic redefinition and the additional non-local tails which would arise. The problem of finding a \emph{non-perturbative} redefinition which relates the above tails to standard HS equations is so far open. In this note we restrict the attention to the lowest non-trivial order.

\item Notice that choosing a different coupling constant for the canonical current amounts to a subleading contribution in \eqref{genred} with spin-dependent behaviour $\sim 1/l^{2s+1}$ for the coefficients $a_l$ (recall that in the minimal type A theory we restrict the attention to even spins $s>0$). Changing the overall coefficient $\alpha_s$ of the current in \eqref{caneq1} by $\epsilon$ \emph{does not} change the leading asymptotic behaviour $\sim 1/l^3$ of the series expansion of the redefinition \eqref{genred}. This implies that the specification of an asymptotic behaviour for the coefficients $a_l$ is \emph{not} sufficient to specify a proper functional class beyond the proposal of \cite{Taronna:2016ats}. Allowing redefinitions whose asymptotic behaviour is $a_l\sim 1/l^3$ does \emph{not} fix a unique value for $\alpha_s$. Some further condition on the redefinitions must be introduced in order recover a unique admissible choice for $\alpha_s$ when enlarging the functional space beyond the proposal of \cite{Taronna:2016ats}.

Notice that the above analysis of the coefficients has a simple interpretation. Given a certain field redefinition with coefficients $a_l$ bound to have a certain asymptotic behaviour, the corresponding improvement can be obtained by a simple action of the covariant adjoint derivative whose effect is to produce some other pseudo-local tail with coefficients $\tilde a_l$ which can be expressed linearly in terms of the coefficients $a_l$, $a_{l-1}$ and $a_{l-2}$:
\begin{equation}
\tilde a_l= A^{(s)}_l\,a_l+B^{(s)}_{l}\,a_{l-1}+C^{(s)}_{l}\,a_{l-2}\,,
\end{equation}
with
\begin{align}
    A^{(s)}_l&\sim (l+s+2)^2\,,& B^{(s)}_l&\sim 2ls+2(l+1)^2+s^2\,,& C^{(s)}_l\sim l^2
\end{align}
This means that $\tilde a_l \prec l^2 a_l$. On the other hand it is \emph{impossible} to obtain coefficients $\tilde a_l$ growing for $l\to\infty$ much more slowly than the original set of coefficients $a_l$. The only possibility is to have some fine-tuning so that the coefficients $\tilde a_l$ go much faster to zero than the original coefficients. This implies that in order to remove by a field redefinition a backreaction with a given asymptotic behaviour for its coefficients the best one can do is to have a redefinition with similar asymptotic behaviour $\tilde a_l \sim a_l$. This simple argument indirectly implies that the redefinition proposed in \cite{Vasiliev:2016xui} should also be compatible with the \emph{spin-independent} asymptotic behaviour presented in this note. To conclude, a simple test of the functional class proposal of \cite{Vasiliev:2015wma,Vasiliev:2016xui} would be to check if the redefinitions \eqref{genspin} for different values of $\alpha_s$ than \eqref{recon} are indeed not admissible.

\item It might be of some interest also to consider a different perspective on the same problem. It is indeed possible to avoid to talk about the subtle issue of field redefinitions and study the limit of the finite derivative truncations of a given back-reaction:
\begin{equation}\label{Jsum}
    J=\sum_{l=0}^\infty \frac{j_l}{l!l!}\,\Box^l\left(\nabla\ldots\nabla\phi\ \nabla\ldots\nabla\phi\right)\equiv \lim_{k\to\infty}\underbrace{\sum_{l=0}^k \frac{j_l}{l!l!}\,\Box^l\left(\nabla\ldots\nabla\phi\ \nabla\ldots\nabla\phi\right)}_{J_k}\,.
\end{equation}
In the above procedure each finite-derivative truncation is well-defined and one can extract the canonical-current piece of each truncation unambiguously. Analogously, one can compute for each truncation the corresponding Witten diagram using standard techniques from local field theories and take the limit only afterwards \cite{Skvortsov:2015lja,Taronna:2016ats}.

We declare that the limit exists when the limit is finite and is \emph{independent} of local redefinitions $f_k$ or $g_k$ performed on each given truncation under the assumption that $f_k$ and $g_k$ converge to admissible redefinitions $f_\infty$ and $g_\infty$ according to\footnote{We briefly recall that in order to check whether a redefinition $f_\infty$  belongs to the functional class of \cite{Taronna:2016ats} one first needs to compute the associated improvement $J^{(f)}$ generated by the field redefinition at this order. The corresponding redefinition is then considered admissible iff the limit of the projections of each local truncation of $J^{(f)}$ on the local canonical coupling is vanishing:
\begin{align}
\lim_{k\to\infty}J_k^{(f)}&=(\lim_{k\to\infty} a_k)\,\mathbf{J}+\lim_{k\to\infty}\Delta J_k\,,& \lim_{k\to\infty} a_k&=0\,.
\end{align}
Here $\mathbf{J}$ is a fixed, but otherwise arbitrary, (\emph{local}) representative for the non-trivial canonical coupling.} \cite{Taronna:2016ats}. Using a diagrammatic language, the existence of the limit can be summarised by the following commutative diagram:
\be
\begin{matrix}\includegraphics[width=10cm,keepaspectratio]{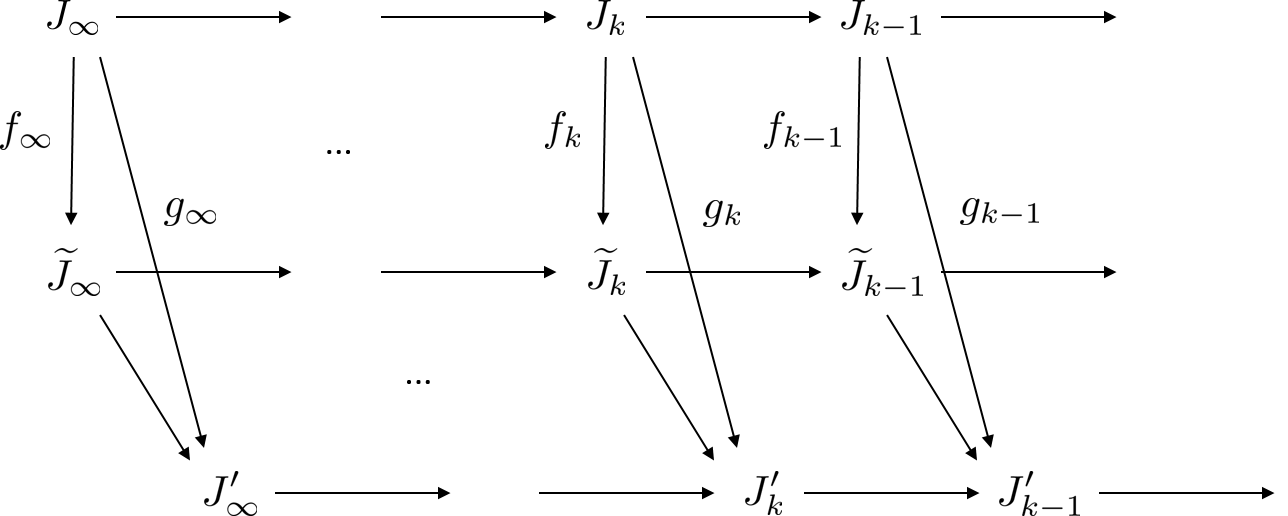}\end{matrix}\,,
\ee
where $f_\infty$ and $g_\infty$ belong to the functional class defined in \cite{Taronna:2016ats} while $\widetilde{J}_k$ and $J^\prime_k$ are different local forms of the truncation which differ by a \emph{local} field redefinition.

If this limit exists we can resum the higher-derivative tail and extract the coefficient of the canonical Metsaev-like coupling. In \cite{Skvortsov:2015lja} it was observed that the above limit for the backreaction \eqref{back} does \emph{not} exist. In this case, it might still be possible to define the sum via some analytic continuation. This is a standard situation where one can define the sum of infinite series formally introducing a cut off procedure. An example of this procedure is:
\begin{equation}
\sum_{l=1}^\infty 1\,,
\end{equation}
which can be regularised introducing a regulator as:
\begin{equation}
\sum_{l=1}^\infty e^{-\epsilon (l+\Lambda)}=\frac{e^{-\epsilon \Lambda}}{e^{\epsilon }-1}\sim \frac{1}{\epsilon}-\left(\frac{1}{2}+\Lambda\right)+O\left(\epsilon\right)\,.
\end{equation}
The choice $\Lambda=0$ reproduces the standard $\zeta$-function regularisation.
As expected, the finite part of the result is regulator dependent and hence ambiguous. For a backreaction \eqref{Jsum} with a divergent sum one similarly ends up with expressions which can be defined formally by analytic continuation. For each given spin there exist a choice of regulator which reproduces the result expected by the holographic reconstruction \eqref{recon}. The question then becomes the same which is usually asked about a renormalisable theory. Namely, whether the choice of regularisation which gives results compatible with holographic reconstruction \eqref{recon} is spin-dependent. If the choice of regulator is spin \emph{independent}, the choice for spin $2$ will fix at the same time the whole backreaction unambiguously. However, if the proper choice of regulator compatible with the holographic reconstruction is spin-dependent the corresponding analytic continuation is not predictive. In the following we give the regularised results for the backreaction for $s=2,4,6$ using the results of \cite{Skvortsov:2015lja,Taronna:2016ats}:
\begin{align}
\alpha_2(\Lambda)&=\frac{1}{36} \left(1-6 \Lambda ^2\right)\,,\\
\alpha_4(\Lambda)&=\frac{-2100 \Lambda ^6+14280 \Lambda ^5-31290 \Lambda ^4+26600 \Lambda ^3-1680 \Lambda ^2+10080 \Lambda +34567}{1058400}\,,\\
\alpha_6(\Lambda)&=\frac{1}{92207808000}\Big(-291060 \Lambda ^{10}+6338640 \Lambda ^9-57387330 \Lambda ^8+280637280 \Lambda ^7\nonumber\\&-802849740 \Lambda ^6+1351860048 \Lambda ^5-1257850440 \Lambda ^4+525866880 \Lambda ^3\nonumber\\&-3991680 \Lambda ^2+79833600 \Lambda +415046341\Big)\,.
\end{align}
It is now easy to verify that the choice of the regulator which matches the holographically reconstructed result \eqref{recon}, in the appropriate normalisation, for $\alpha_2$ is not consistent with $\alpha_4$ or $\alpha_6$ with the same normalisation. Therefore, the regulator $\Lambda$ must be spin-dependent, to compensate the highly spin-dependent form of the above regularised expressions. This makes the corresponding analytic continuations unpredictive.

This result is not in contradiction with the analysis of field redefinitions presented in this note. The above feature may be a different reincarnation of the fact that all redefinitions removing the higher-derivative tail, regardless the value of $\alpha_s$, have the same asymptotic behaviour at $l\to\infty$. Similarly, this does not allow to single out a unique value of $\alpha_s$.

\item It might be interesting to compare the complicated redefinition \eqref{genred} which matches the holographically reconstructed result starting from \eqref{back} with other redefinitions which would generate the required coupling constant but from a free theory. One may indeed start with \emph{free} Fronsdal equations and find the non-local redefinition which would generate the appropriate cubic couplings. This redefinition should not be admissible but, remarkably, it has a faster asymptotic behaviour for $l\to \infty$ than \eqref{genred}:
\begin{equation}\label{genredfree}
a_{l-1}=\alpha_s\,s!(s-1)!\,(2l+s+2)\left(\frac{l!}{(l+s+1)!}\right)^2\,.
\end{equation}
This expression is simpler than \eqref{genred}, and falls off faster as $l\to\infty$: $a_l\sim\frac{1}{l^{2s+1}}$\,. Notice that the above redefinition \eqref{genredfree} allows to generate the holographic backreaction from the free theory with the choice \eqref{recon}. The above redefinition however should not be considered admissible as it does not leave the cubic Witten diagrams invariant.

\item In the parity violating case, the backreaction \eqref{back} is multiplied by a factor proportional to the parity violating phase $\theta$. Surprisingly (see \cite{Kristiansson:2003xx,Boulanger:2015ova}) this factor is given by $\cos(2\theta)$:
\begin{equation}\label{back2}
J_{\mu(s)}(\theta)=\sum_{l=0}^\infty\,\frac{j_l\,\cos(2\theta)}{l!l!}\,\Box^l\left(\nabla\ldots\nabla\phi\ \nabla\ldots\nabla\phi\right)_{\mu(s)}\,,
\end{equation}
It was then observed in \cite{Boulanger:2015ova} that term by term each element of the pseudo-local series in \eqref{back} vanishes identically for $\theta=\frac{\pi}{4}$. The interpretation of this $\theta$-dependence is at the moment unclear as it seems to be in contradiction with the holographic expectations \cite{Maldacena:2012sf}. In \cite{Vasiliev:2016xui} it was proposed that up to an admissible field redefinition one is left with a canonical current and a $\theta$-independent coefficient.
Let us assume that the field redefinition reproducing the holographically expected coupling constants is admissible. This means that the difference between the above non-local current \eqref{back2} and a local canonical current with a fixed non-vanishing coefficient\footnote{Whether or not $\alpha_s(\theta)$ depends on $\theta$ is not important for the following argument. The only assumption is that $\alpha_s$ does not vanish for any value of $\theta$.} $\alpha_s(\theta)$ is an admissible improvement:
\begin{equation}
    J_s(\theta)-\alpha_s(\theta) J_s^{\,\text{can.}}=\Delta^{(J)}_s(\theta)\,.
\end{equation}
Since the above must be true for any value of $\theta$, we can now set $\theta=\tfrac{\pi}{4}$ and use that $J_s(\tfrac{\pi}{4})=0$. This however implies that the canonical current itself is an admissible improvement:
\begin{equation}
-\alpha_s(\theta) J_s^{\text{can.}}=\Delta^{(J)}_s(\tfrac{\pi}{4})\,,
\end{equation}
This further implies that $\Delta^{(J)}_s(\theta)+\frac{x}{\alpha_s(\theta)}\,\Delta^{(J)}_s(\tfrac{\pi}{4})$ is an admissible improvement whose associated admissible field redefinition puts \emph{any} coefficient (parametrised by $x$) in front of the canonical current. However, this contradicts our assumption on the admissibility of the above field-redefinition.
Furthermore, the above shows that in the $\theta=\frac{\pi}{4}$ case the initial redefinition itself must generate the canonical backreaction \emph{from} the free Fronsdal equations, and therefore has to match \eqref{genredfree}. Let us also stress that the only assumptions we used is the existence of a field redefinition which is both allowed and capable of changing the $\theta$ dependence of the initial pseudo-local current \eqref{back2}. The contradiction we find seems to imply that no such admissible redefinition may exist.
\end{enumerate}
In the following we give some details on the results summarised above. This note is organised as follows. After a brief review of the main formalism in Section~\ref{review_part}, we move to the analysis of the redefinitions and to the study of their structure in Section~\ref{Redefs}. We describe the analytic continuations of the higher-derivative tails in Section~\ref{analytic}. We end with a short summary and some outlook in Section~\ref{Conclusions}. In the Appendix we summarise various conventions and derive the normalisation of the Fronsdal kinetic term used in the unfolded language.

\section{Pseudo-local Currents}\label{review_part}
In \cite{Prokushkin:1999xq,Kessel:2015kna,Boulanger:2015ova,Skvortsov:2015lja,Taronna:2016ats} a convenient generating function formalism was developed, first in 3d and then in 4d, to manipulate quasi-local current interactions and corresponding field redefinitions. In this section we recall the basic ingredients of the formalism, and refer to \cite{Kessel:2015kna,Skvortsov:2015lja} for further details. The main object is the zero-form $C(y,\bar{y}|x)$, which is a formal expansion in the spinorial oscillators $y_\ga$ and $\bar{y}_\gad$ satisfying the linearised unfolded equations
\begin{equation}
\tadD C(y,\bry|x)=0\,.\label{twisteq}
\end{equation}
Here $\adD$ and $\tadD$ are the adjoint and twisted-adjoint covariant derivatives expressed in terms of spinorial oscillators as:\footnote{$h^{\ga\gad}$ is the vierbein, $\nabla=d -\omega^{\ga\ga}y_\ga\pl_\ga-\varpi^{\gad\gad} \bry_\gad\bar{\pl}_\gad$ is the AdS$_4$ covariant derivative, while $\varpi^{\ga\ga}$ and $\varpi^{\gad\gad}$ are the (anti-)self-dual components of the spin-connection of AdS$_4$.}
\begin{align}
\adD &=\nabla-h^{\ga\gad}(y_\ga \bar{\pl}_\gad-\bry_\gad {\pl}_\ga)\,,\\
\tadD &=\nabla+ih^{\ga\gad}(y_\ga \bry_\gad-\pl_\ga \bar{\pl}_\gad)\,.
\end{align}
In the unfolded language, the zero-form is the main ingredient upon which one constructs ordinary current interactions. Furthermore, upon solving the twisted adjoint covariant constancy condition one recovers the relations between components of the zero form and derivatives of the linearised Weyl tensors associated to the HS fields:
\begin{align}
C_{\ga(k+2s),\gad(k)}(x)&\sim\nabla^k C_{\ga(2s)}(x)\,,& C_{\ga(k),\gad(k+2s)}(x)&\sim\nabla^k C_{\gad(2s)}(x)\,,
\end{align}
Above, $C_{\ga(2s)}(x)$ and $C_{\gad(2s)}(x)$ are the self-dual and anti-self dual part of the linearised HS Weyl tensor for $s>0$ and $\Phi(x)\equiv C(0,0|x)$ is the actual scalar field.

A current which sources the HS Fronsdal operator is a bilinear functional $J_{\ga(s)\gad(s)}(C,C)$ of the zero-form $C$, which is conserved on the equations of motion \eqref{twisteq}:
\begin{align}
\nabla^{\gb\gbd}J_{\gb\ga(s-1)\gbd\gad(s)}\approx0\,.
\end{align}

For practical purposes, it is convenient to specify the most general form for a current in terms of a generating function Kernel $\bar{J}(Y,\xi,\eta)$:
\begin{equation}
J(C,C)= \int  d^4\xi \, d^4\eta\, \bar{J}(Y,\xi,\eta)\,C(\xi|x) C(\eta|x)\,.
\end{equation}
Above we have introduced the Fourier transform of the $0$-form with respect to the spinorial variables $y$ and $\bry$,
\begin{equation}
C(y,\bry|x)=\int d^2\xi d^2\brxi\,e^{iy^\ga \xi_\ga+i\bry^\gad\brxi_{\gad}}\,C(\xi,\brxi|x)\,,
\end{equation}
where for ease of notation we denote both Fourier transforms and original $0$-forms by the same letter.
In this way, contractions of indices are encoded as simple monomials:
\begin{align}
(y\xi)^n(y\eta)^m(\bry\brxi)^{\bar{n}}(\bry\breta)^{\bar{m}}(\xi\eta)^l(\brxi\breta)^{\bar{l}}\sim C_{\ga(n)\nu(l);\gad(\bar{n})\dot\nu(\bar{l})}C_{\ga(m)}{}^{\nu(l)}{}_{\gad(m)}{}^{\dot{\nu}(\bar{l})}\,.
\end{align}

In addition to the above way of representing a generic current interaction it is also convenient to introduce generating functions of coefficients via contour integrals, as originally proposed in \cite{Prokushkin:1999xq}. Restricting the attention to the canonical current sector, which in four-dimensions is uniquely specified by the absence of trace components, one can then write the most general current kernel as
\begin{align}\label{can4d}
\bar{J}(Y,\xi,\eta)=\oint_{\tau_i,s,r} j(\alpha_1,\alpha_2,\beta,\gamma) \, e^{i(s y\zeta^-+\tau_1\xi\eta+r \bry\brzeta^++\tau_2\brxi\breta)}\,,
\end{align}
with contour integrals in $\tau_i$, $s$ and $r$ and in terms of a function of $4$ complex variables:
\begin{align}
\alpha_1&=\tau_1^{-1}\,,& \alpha_2&=\tau_2^{-1}\,,&\beta&=s^{-1}\,,&\gamma&=r^{-1}\,,\\
&&\zeta^{\pm}&=\xi\pm\eta\,,&\brzeta^{\pm}&=\brxi\pm\breta\,.
\end{align}
These parametrise the four contractions of indices relevant to the canonical current sector in 4d. One can easily translate between the contour integral form and the generating function form via:
\begin{equation}
\alpha_2^{m+1} \alpha_1^{n+1} \beta ^{s_1+1} \gamma ^{s_2+1}\rightarrow \frac{(i\brxi\breta){}^{m} (i\xi\eta){}^{n} (i\bry\brzeta^+){}^{s_2} (iy\zeta^-){}^{s_1}}{m!\, n!\, s_1!\, s_2!}\,.
\end{equation}
Notice that in this generating function calculus the function $j(\alpha_1,\alpha_2,\beta,\gamma)$ is defined as a formal series, and this should be thought at this level as a regularity condition. This regularity condition will be assumed here since it is equivalent to pseudo-locality of the back-reaction. Notice also that a constant term or terms proportional to $1/\alpha_1$ drop out of the contour integration. In particular, two generating functions will give the same pseudo-local current if they differ by terms of this type. For the details we refer to \cite{Prokushkin:1999xq,Kessel:2015kna} and use the symbol $\sim$ to indicate equality modulo the above equivalence relation.

To conclude this section we present the corresponding expressions for the explicit Fronsdal currents of the type $s$-$s_1$-$s_2$:
\begin{equation}\label{Fr1}
\frac{1}{4(s-1)}\underbrace{\left[\Box+2(s^2-2s-2)+\ldots\right]\phi_s}_{\mathcal{F}}=J_s(C,C)\,,
\end{equation}
as extracted from Vasiliev's equations\footnote{The normalisation for the Fronsdal tensor comes from the solution to torsion as described in \cite{Boulanger:2015ova}.} in \cite{Boulanger:2015ova}. The generating function $j_s$ for the spin-s current is:
\begin{equation}\label{Res}
j_s(\beta,\gamma,\alpha_1,\alpha_2)=\frac{i}2e^{2i\theta}(\beta\gamma)^{s+1}\sum_{n,m}c_{n,m}^{(s)}\left(\frac{ p_1(n)-m\, p_2(n)}{(n+s-1)^2 (n+s)^2 (n+s+1)}\right)\alpha_1^n\alpha_2^m+h.c.\,,
\end{equation}
to be plugged into \eqref{can4d} with
\begin{align}
p_1^{(s)}(n)&=\Big[[2(s-1)]n^3+[2+s (6 s-5)]n^2+[s \left(1-4 s+6 s^2\right)]n\nonumber\\&+[s (1+s) (2+s (2 s-3))]\Big]\,,\\
p_2^{(s)}(n)&=\Big[[2(s-1)]n^2+[2+s (4 s-3)]n+s (1+s (2 s-1))\Big]\,,
\end{align}
and
\begin{align}
c_{n,m}^{(s)}=\frac{1}{4(s-1)}\frac{(-1)^{m+s}+(-1)^{n}}{2}\,.
\end{align}

Notice that powers of $\alpha_1$ and $\alpha_2$ translate into contractions among the $0$-form, hence powers of $\ga_1\ga_2$ parametrise powers of $\Box$ in the metric-like language. We can then define $\tau=\ga_1\ga_2$ parametrising the pseudo-local tail of the given interaction term.
In general, when restricting attention to sources to the Fronsdal equation in the spin-s sector, one fixes the dependence on $\beta$ and $\gamma$ as in \eqref{Res} while working with a generating function of the type:
\begin{align}\label{Generic}
j_{s}(\beta,\gamma,\alpha_1,\alpha_2)&=i \cos(2\theta)(\beta\gamma)^{s+1}\,\alpha_1^n\,g(z)\,,& z&=\alpha_1\alpha_2\,.
\end{align}
The function $g(z)$ parametrises the infinite non-local tail while the dependence on $\beta$, $\gamma$ and a single $\alpha_1$ ($\alpha_2$) fixes the canonical current tensor structure and the spin of the zero-forms (see Appendix \ref{Dictionary}). The canonical current with no higher-derivative tail $J^{\text{can}}(C,C)$ is simply encoded by the choice $g(z)\sim \alpha\, z$.

In the $s$-$0$-$0$ case the dictionary can be given quite explicitly as:
\begin{align}\label{dic}
j_s&\rightarrow \cos(2\theta)\sum_{l,k}a_{l,k}\nabla_{\mu(s-k)\nu(l)}\Phi\nabla_{\mu(k)}{}^{\nu(l)}\Phi\,,\\
\nabla_{\mu(s-k)\nu(l)}\Phi\nabla_{\mu(k)}{}^{\nu(l)}\Phi&\equiv h^{\ga\gad}_{\mu}\ldots h^{\ga\gad}_{\mu}C_{\ga(s-k)\nu(l),\gad(s-k)\dot{\nu}(l)}C_{\ga(k)}{}^{\nu(l)}{}_{,\gad(s-k)}{}^{\dot{\nu}(l)}\,,\label{def}
\end{align}
with
\begin{align}\label{factor}
a_{l,k}&=\frac{(-1)^{k+l}s!s!}{l!l!k!k!(s-k)!(s-k)!}\,g_l\,,& g(z)=\sum_{l=0}^\infty g_{l}\,z^{l+1}\,.
\end{align}
The above dictionary holds also for redefinitions of the Fronsdal field and allows to easily translate from the generating function language to the standard tensorial language. Notice that the notation $\nabla_{\mu(s-k)\nu(l)}\Phi\nabla_{\mu(k)}{}^{\nu(l)}\Phi$ is defined by eq.~\eqref{def} and includes symmetrisation and traceless projection. In particular $\nabla_{\mu(s)}\neq \nabla_\mu\ldots \nabla_\mu$. We give further details on more general current interactions included in the generating function \eqref{Generic} in Appendix~\ref{Dictionary}.

In the following we shall restrict our attention to the function $g(z)$ encoding the higher-derivative tail. It is important to stress that the $\frac1{(l!)^2}$ factor in \eqref{factor} arises via the above contour integrations and does not appear in the function $g(z)$.

\section{Pseudo-Local Field Redefinitions}\label{Redefs}

In this section we employ the generating function formalism to study the non-local field redefinitions which relate the back-reaction extracted from Vasiliev's equation in \cite{Boulanger:2015ova} to canonical currents. For simplicity we work in the A-type model, setting $\theta=0$. The discussion generalises straightforwardly to any choice of $\theta$ as this only appears as an overall factor. The effect of a field redefinition of the spin-$s$ field quadratic in the $0$-form can be encoded in an arbitrary function $k(z)$ (analogous to $g(z)$ above) via the differential operator\footnote{The corresponding redefinition at the level of unfolded equations reads:
\begin{align}
\delta\omega_s(y,\bry)&=h^{\ga\gad}\pl_\ga\bar{\pl}_\gad k_s(y,\bry)\,.
\end{align}}
\begin{multline}\label{canonical}
\Delta j_s=\frac{(\beta\gamma)^{s+1}}{4(s-1)}\Big[z^2  (z +1)^2 k_s''(z)+z(z+1)(2 s+(3+\bar{n}) z +3+\bar{n}) k_s'(z)\\+ \left(\bar{n}(z+1)(z+s+1)+s^2 (z +1)+2 s+(z +1)^2\right)k_s(z)\Big]\,,
\end{multline}
where $\bar{n}=n-m$ with $n$ and $m$ giving the power of $\alpha_1$ and $\alpha_2$ respectively as in \eqref{Res}.
In this way, the problem of finding redefinitions which remove the pseudo-local tail is thus mapped into a ordinary differential equation.

In the following we apply this formalism to the current \eqref{Res}. We first give the explicit field redefinition which maps the pseudo-local current \eqref{Res} to a canonical current with an arbitrary overall coefficient:
\begin{equation}\label{caneq}
\underbrace{\Box\phi_s+\ldots}_{\mathcal{F}}=\alpha_s \,J_s^{\text{can}}(C,C)\,.
\end{equation}
Notice that in principle $\alpha_s$ can depend on the spins $s_1$ and $s_2$ of the zero-forms. However, since the result of the redefinition depends only on $s$ we do not write explicitly the dependence on $s_1$ and $s_2$.
We then compare redefinitions which give different overall coefficients $\alpha_s$ for the canonical current. For ease of notation we only restrict ourselves to the currents with $\bar{n}=0$ (see Appendix \ref{Dictionary}).

\paragraph{Spin-2:}
Using the generating function formalism reviewed above, the back-reaction \eqref{Res} for $s=2$ and $\bar{n}=0$ can be encoded by (see eq.~\eqref{Generic}):
\begin{equation}
g_2(z)=\sum_{l=0}^\infty \frac{1}{2\, (l+1)^2 (l+2)}\,(-z)^l=\frac{z -z\,\text{Li}_2(-z )-(z +1) \log (z +1)}{2 z ^2}\,.
\end{equation}
The field redefinition that reduces the above pseudo-local current to its canonical form can be obtained solving the differential equation:
\begin{align}
z  (z +1) \left(z  (z +1) k_2''(z)+(3 z +7) k_2'(z )\right)+(z +3)^2 k_2(z )=-4\,g_2(z)+\alpha_2\, z+\beta_2\,.
\end{align}
The constant $\beta_2$ parametrises non-trivial pseudo-local redefinitions which result in a vanishing\footnote{We have not considered terms on the right-hand side of the type $z^{-m}$, since the corresponding redefinitions vanish upon performing the contour integration and are thus equivalent under the equivalence relation $\sim$. The inequivalent solutions are parametrised by one constant $\beta_s$. This can be checked by solving the differential equation (see \cite{Kessel:2015kna} for further details).} contribution to the backreaction upon performing the contour integration. On the other hand, $\alpha_2$ is an arbitrary constant in front of the canonical current (which in this case is the stress tensor \eqref{caneq}).
In the following we fix the constant $\beta_2=2\alpha_2 +2$ in such a way that the slowest contribution to the coefficients in the $l\to\infty$ limit is set to zero. Furthermore, one should carefully fix the freedom in the homogeneous solutions choosing the unique solution to the above differential equation which is analytic in $z$, as this corresponds to the standard regularity condition on pseudo-local functionals \cite{Prokushkin:1999xq,Skvortsov:2015lja}. The final result has the following series expansion around $z=0$:
\begin{equation}
k_2(z)=\sum_{l=0}^\infty\frac{(4 \alpha_2 +l (l+3)+2)}{(l+1)^2 (l+2) (l+3)^2}\,(-z)^l\,,
\end{equation}
where we have left $\alpha_2$ arbitrary. So far the discussion applies to any current interaction with $\bar{n}=0$ regardless the spin of the 0-forms. Considering the $2$-$0$-$0$ case, the choice compatible with holography is given by \eqref{recon} and is $\alpha_2=2 g^2\,N_0^2$.

\paragraph{Spin-4:}
The source to the spin-4 Fronsdal operator extracted from \eqref{Res} for $\bar{n}=0$ can be encoded in the following generating-function:
\begin{equation}
g_4(z)=\sum_{n=0}^\infty \frac{(3 l+11)}{3 (l+3)^2 (l+4)^2}\,(-z)^l\sim-\frac1{3\,z^4}\,\left[(2 z+1) \text{Li}_2(-z) + (z+1) \log (z+1)\right]\,.
\end{equation}
The redefinition bringing the above back-reaction to its canonical form \eqref{caneq} with an arbitrary coefficient $\alpha_4$, can be obtained solving the differential equation:
\begin{multline}
z  (z +1) \left(z  (z +1) k_4''(z)+(3 z +11) k_4'(z )\right)+(z^2+18z+25) k_4(z )\\=-12\,g_4(z)+\alpha_4 z+\beta_4\,.
\end{multline}
Again, $\beta_4$ parametrises pseudo-local redefinitions which do not change the back-reaction, and can be chosen to improve the asymptotic behaviour. In this case we choose $\beta_4=\frac{1}{2} (3 \alpha_4 +4)$.
The solution to the above differential equation is then:
\begin{equation}
k_4(z)=\sum_{l=0}^\infty\frac25\frac{p_4(l)+120 (6 \alpha_4 +5)}{(l+1)^2 (l+2)^2 (l+3) (l+4)^2 (l+5)^2}\,(-z)^l\,,
\end{equation}
with $p_4(l)$ a polynomial of order $5$ in the variable $l$:
\begin{equation}
p_4(l)=l \left(5 l^5+77 l^4+470 l^3+1445 l^2+2345 l+1898\right)\,.
\end{equation}

\paragraph{Spin-6:}

The source to the spin-6 Fronsdal operator extracted from \eqref{Res} for $\bar{n}=0$ can be encoded in the generating-function
\begin{multline}
g_6(z)=\sum_{l=0}^\infty \frac{3 (5 l+28)}{10 (l+5)^2 (l+6)^2}\,(-z)^l\\\sim-\frac3{10\,z^6}\left[(z +1) \log (z +1)+ (3 z +2) \text{Li}_2(-z )\right]\,.
\end{multline}
The redefinition bringing the above back-reaction to its canonical form \eqref{caneq} with an arbitrary coefficient $\alpha_6$ can be obtained solving the differential equation:
\begin{multline}
z  (z +1) \left(z  (z +1) k_6''(z)+3(z +5) k_6'(z )\right)+z  (z +38)+49 k_6(z )\\=-20\,g_6(z)+\alpha_6 z+\beta_6\,.
\end{multline}
Again, $\beta_6$ parametrises pseudo-local redefinitions which do not change the back-reaction. In this case we choose $\beta_6=\frac{1}{3} (4\, \alpha_6 +6)$.
The solution to the above differential equation is then given by:
\begin{equation}
k_6(z)=\sum_{l=0}^\infty\frac{p_6(l)+2\cdot 6!\cdot 5!\, (\alpha_6 +1)}{(l+1)^2 (l+2)^2 (l+3)^2 (l+4) (l+5)^2 (l+6)^2 (l+7)^2}\,(-z)^l\,,
\end{equation}
with $p_6(l)$ a polynomial of order $10$ in the variable $l$:
\begin{multline}
p_6(l)=l\, \Big(3 l^9+\frac{323 l^8}{3}+1686 l^7+\frac{105862 l^6}{7}+85755 l^5+320009 l^4\\+792684 l^3+\frac{3844612 l^2}{3}+1288512 l+\frac{5072640}{7}\Big)\,.
\end{multline}

\paragraph{Generic even spin:}

In the generic spin-$s$ case, the source quadratic in the $0$-form to the Fronsdal operator extracted in \cite{Boulanger:2015ova} can be encoded by the generating-function
\begin{multline}
g_s(z)=\sum_{l=0}^\infty \frac{s (2 l (s-1)+s (2 s-3)+2)}{8 (s-1) (l+s-1)^2 (l+s)^2}\,(-z)^l\\\sim-\frac{c_s}{z^s}\left[(z +1) \log (z +1)+\tfrac12 (s(z+1)-2) \text{Li}_2(-z )\right]\,,
\end{multline}
with
\begin{equation}
c_s=\frac14\frac{s}{s-1}\,.
\end{equation}
The corresponding redefinition mapping the above pseudo-local back-reaction to a canonical current \eqref{caneq} with overall coefficient $\alpha_s$ \eqref{caneq} is:
\begin{equation}\label{genspin}
k_s(z)=\sum_{l=0}^\infty\frac{2l+s+2}{2}\left(\frac{l!}{(l+s+1)!}\right)^2[p_s(l)+X_s\, \alpha_s+Y_s]\,(-z)^l\,,
\end{equation}
with $p_s(l)$ a polynomial of degree $2(s-1)$ in the variable $l$ and $X_s$ and $Y_s$ two spin-dependent constant. The general form of the polynomial for arbitrary spin is cumbersome, and we do not present it explicitly.

To summarise, in this section we have calculated explicitly the field redefinition relating the backreaction extracted from Vasiliev's equations to their canonical form. The main result is that the coefficient of the canonical current turns out to contribute a subleading term in the field redefinition and cannot be fixed by prescribing an asymptotic behaviour for the corresponding coefficients.

\section{Analytic Continuation}\label{analytic}

In this section upon reviewing the results of \cite{Skvortsov:2015lja,Taronna:2016ats}, we study the analytic continuation of the formal series obtained by considering the limit of the finite derivative truncations of the backreaction \eqref{back}.

The main observation of \cite{Skvortsov:2015lja,Taronna:2016ats} is that for any truncation of the pseudo-local interaction term \eqref{Res} it is possible to compute the corresponding projection on the canonical current piece which is parametrised by finitely many structures in correspondence with Metsaev's cubic couplings. Each higher-derivative term gives a contribution to the canonical coupling weighted by some proportionality factor $C_l^{(s)}$  which measures the overlap of the higher-derivative term on the canonical structure. Schematically the projection reads:
\begin{equation}
\mathcal{P}\left[\Box^l\left(\nabla\ldots\nabla\phi\ \nabla\ldots\nabla\phi\right)_{\mu(s)}\right]=C_{l}^{(s)}\,\left(\nabla\ldots\nabla\phi\ \nabla\ldots\nabla\phi\right)_{\mu(s)}\,,
\end{equation}
where indeed the right-hand side is proportional to the canonical structure times a certain overall coefficient which was computed in \cite{Skvortsov:2015lja,Taronna:2016ats}. In generating functions terms, the above amounts to the following projection (we restrict again to $\bar{n}=0$):
\begin{equation}
    \mathcal{P}[z^l]=C_l^{(s)}\,z\,,
\end{equation}
with coefficients given by:
\begin{equation}
C_l^{(s)}=-\frac{(-1)^l s\, \Gamma (l+s+1) \, _3F_2(1-s,1-s,-2 s;2-2 s,l-s+1;1)}{2 (2 s-1) \Gamma (s+1)^2 \Gamma (l-s+1)}\,.
\end{equation}

After the projection the canonical structure $g(z)=z$ factorises and one is left with an overall coefficient which combines together all contributions from each higher-derivative term. Below we give the corresponding coefficient for some low spin examples:
\begin{align}
s&=2\,,& &-\frac1{12}\sum_{l=1}^\infty l\,,\\
s&=4\,,& &-\frac1{3\cdot 7!}\sum_{l=1}^\infty \frac{l (l+1) (l+2)^2 (3 l+11) (5 l (l+4)+3)}{(l+3) (l+4)}\,,\\
s&=6\,,& &-\frac{3}{5\cdot 11!}\sum_{l=1}^\infty \frac{(l+4)!}{(l-1)!}\frac{(l+3) (5 l+28) (7 l (l+6) (3 l (l+6)+19)+20)}{(l+5) (l+6)}\,,\\
s&=8\,,& &-\frac{7}{6\cdot 15!}\sum_{l=1}^\infty \frac{(l+6)!}{(l-1)!}\tfrac{(l+4) (7 l+53) (l (l+8) (11 l (l+8) (13 l (l+8)+274)+14631)+420)}{ (l+7) (l+8)}\,.
\end{align}
These series are divergent and can be regularised by a standard $\zeta$-function regularisation introducing a regulator of the type $e^{-\epsilon(l+\Lambda)}$ which allows to resum them. Dropping the divergent parts and setting $\epsilon$ to zero one then obtains the following regularised expressions for the corresponding overall coefficients of the canonical current:\footnote{In the formulas for $\alpha_s(\Lambda)$ we take into account the normalisation $\frac1{4(s-1)}$ in \eqref{Fr1} so to arrive to a source of the type $\mathcal{F}_{\mu(s)}=\alpha_s(\Lambda) J_{\mu(s)}^{\text{can.}}$.}
\begin{align}
\alpha_2(\Lambda)&=\frac{1}{36} \left(1-6 \Lambda ^2\right)\\
\alpha_4(\Lambda)&=\frac{-2100 \Lambda ^6+14280 \Lambda ^5-31290 \Lambda ^4+26600 \Lambda ^3-1680 \Lambda ^2+10080 \Lambda +34567}{1058400}\\
\alpha_6(\Lambda)&=\frac{1}{92207808000}\Big(-291060 \Lambda ^{10}+6338640 \Lambda ^9-57387330 \Lambda ^8+280637280 \Lambda ^7\nonumber\\&-802849740 \Lambda ^6+1351860048 \Lambda ^5-1257850440 \Lambda ^4+525866880 \Lambda ^3\nonumber\\&-3991680 \Lambda ^2+79833600 \Lambda +415046341\Big)\\
\alpha_8(\Lambda)&=\frac{1}{39269461271040000}\Big(-51531480 \Lambda ^{14}+2307024720 \Lambda ^{13}-45901996140 \Lambda ^{12}\nonumber\\&+535422167280 \Lambda ^{11}-4065466421016 \Lambda ^{10}+21086692426800 \Lambda ^9\nonumber\\&-76212086580630 \Lambda ^8+191985197049360 \Lambda ^7-330826659683520 \Lambda ^6\nonumber\\&+372883114251648 \Lambda ^5-249836835568320 \Lambda ^4+77398236115200 \Lambda ^3\nonumber\\&-62270208000 \Lambda ^2+2615348736000 \Lambda +16930453296697\Big)\,.
\end{align}
These results allow in principle to fix the regulator in order to recover the holographically reconstructed result of \cite{Bekaert:2015tva,Sleight:2016dba}. The main observation is however that the associated choice for the regulator is \emph{spin-dependent}. Furthermore the regularised coupling constant is given by complicated polynomials of order $\Lambda^{2(s-1)}$. The lack of a spin-independent regularisation makes the corresponding analytic continuations unpredictive as a consequence of the strongly coupled nature of the higher-derivative tails. It is therefore not clear how to identify a regularisation which preserves all HS symmetries without solving the Noether procedure up to the quartic order. On the other hand, the simple form of the local holographically reconstructed couplings \cite{Sleight:2016dba} predicts a very simple structure for the corresponding local interactions.

\section{Conclusions}\label{Conclusions}
In this note we have determined explicitly the general form of the redefinition which reduces the back-reaction extracted in \cite{Boulanger:2015ova} from Vasiliev's equations to a canonical current with an arbitrary overall coefficient \eqref{caneq}. We have also given some details about the analytic continuation of the formally divergent sums. Our analysis provides a convenient test ground to probe functional class proposals for admissible non-local interactions.

The main conclusion of this note is that a full non-perturbative functional class able to fix the overall coefficient in front of the canonical interaction terms should be spelled out to complete the dictionary between Vasiliev's equations and standard HS equations like Fronsdal's equations. Furthermore, it is worth stressing that in this note we have restricted the attention to the cubic order which admits a local completion at least for fixed spins. At the quartic and higher orders we expect the situation to be even more subtle, since non-localities may not be anymore removed by redefinitions, and very restrictive consistency requirements will be put in place by consistency \cite{Taronna:2011kt,Taronna:2012gb,Taronna:2016ats,Rahman:2016tqc}. The study of locality is at the moment incomplete at quartic and higher orders and this complicates the definition and the test of possible non-perturbative functional classes.

We conclude this note with a short list of observations and comments:
\begin{itemize}
\item The main difficulty of HS theories is the absence of a scale beyond the AdS-radius. This implies that the behaviour of a pseudo-local tail can only be controlled by its coefficients owing to $\Lambda[\nabla,\nabla]\sim1$. A functional class proposal would then prescribe a given asymptotic behaviour for the coefficients of higher and higher order terms as in the proposal of \cite{Taronna:2016ats}. Enlarging the functional class beyond the latter proposal seem to lead to unpredictive results unless some further condition on the redefinition is imposed.
\item It would be interesting to study the limit $\alpha^\prime\to\infty$ at the level of effective cubic string field theory couplings in AdS. Some issues about commutativity of limits may arise in this context when taking the limit $\alpha^\prime\to\infty$ before or after removing the higher-derivative tails. It might well be that if one considers the naive $\alpha^\prime\to\infty$ limit of a non-local string coupling this would indeed be strongly coupled. This would mean that the tensionless limit should be well-defined only in a particular field frame, while singular in others. This kind of situation suggests that allowed redefinitions for any finite value of $\alpha^\prime$ could become not allowed after the limit. If so the infinities observed in this note could be resolved upon taking the limit from string theory using $\alpha^\prime$ as regulator for the strongly coupled series. On the other hand, the fact that the $\alpha^\prime\to\infty$ limit could display such subtleties requires particular care in the definition itself of tensionless strings.
\item Beyond the particular problem of mapping the tail in \eqref{back} to a canonical form, the simplicity of the 4d theory in the spinorial language is expected to manifest in a simple form of all quadratic sources when rewritten in the unfolded form. Such rewriting, and the structures involved beyond the $\star$-product, is at the moment unknown (see however \cite{Vasiliev:1988sa,Vasiliev:1989xz,Vasiliev:1989yr}). For this reason an interesting problem would be to unfold the non-linear Fronsdal equations in 4d coming from the cubic couplings extracted holographically in \cite{Sleight:2016dba}. This result would provide us with the complete list of cubic vertices in the unfolded equations possibly giving us a hint of higher-order completions.
\end{itemize}

\section*{Acknowledgments}
\label{sec:Aknowledgements}

I am grateful to N. Boulanger, S. Didenko, D. Francia, A. Sagnotti, E. Sezgin, C. Sleight, Z. Skvortsov, D. Sorokin, P. Sundell and M. Vasiliev for useful discussions. The research of M. Taronna is partially supported by the Fund for Scientific Research-FNRS Belgium, grant FC 6369 and by the Russian Science Foundation grant 14-42-00047 in association with Lebedev Physical Institute. This research was also supported by the Munich Institute for Astro- and Particle Physics (MIAPP) of the DFG cluster of excellence “Origin and Structure of the Universe”.

\begin{appendix}
\section{Canonical Currents Quadratic in the Curvatures}\label{Dictionary}
In this appendix we would like to give a few more details on the explicit tensorial form of the currents discussed in this note. Canonical currents are encoded as generating functions by $(\beta\gamma)^{s+1}\alpha_1^n\,z$, since any dependence on $\alpha_1\alpha_2$ can be removed by a local field redefinition.
Such monomial corresponds to the following generating function kernels:
\begin{multline}
(\beta\gamma)^{s+1}\alpha_1^{n}\,z\rightarrow\frac1{s!^2n!}\,(-i\pl_{\xi}\cdot\pl_{\eta})^n(-iy(\pl_\xi-\pl_\eta))^s(-i\bry(\pl_\brxi+\pl_\breta))^s\,\\\sum_{p_1,q_1}\frac1{p_1!q_1!}C_{\ga(p_1)\gad(q_1)}(x)\xi^{\ga(p_1)}\brxi^{\gad(q_1)}\sum_{p_2,q_2}\frac1{p_2!q_2!}C_{\gb(p_2)\gbd(q_2)}\eta^{\gb(p_2)}\breta^{\gbd(q_2)}\Big|_{\eta=0,\xi=0}\,.
\end{multline}
It is then easy to perform all required differentiations ending up with
\begin{equation}
(\beta\gamma)^{s+1}\alpha_1^{n}\,z\\\rightarrow(-1)^s\,\frac{y^{\ga(s)}\bry^{\gad(s)}}{s!^2}\sum_{p,q=0}^s(-1)^q\binom{s}{p}\binom{s}{q}\,\frac{(-i)^{n}}{n!}C_{\ga(s-p)\gb(n)\gad(s-q)}C_{\ga(p)}{}^{\gb(n)}{}_{\gad(q)}\,,
\end{equation}
together with its conjugate piece which can be obtained by replacing $\alpha_1$ with $\alpha_2$.
Notice that the spin of a zero form $C_{\ga(n)\gad(m)}$ is given by $\tfrac{|n-m|}{2}$ so that the first 0-form has spin $s_1=\tfrac{|n-(p-q)|}2$ while the second has spin $s_2=\tfrac{|n+(p-q)|}{2}$. The number of derivatives is instead $s_1+s_2+\text{min}(s-p+n,s-q)+\text{min}(p+n,q)$. A particularly interesting case is when the canonical current has the maximum number of derivatives $s+s_1+s_2$. For $s_1=s_2$ this coupling is realised with $p=q$ and $s_1=s_2=\tfrac{n}2$. The $s$-$0$-$0$ current is a particular case of the latter for $n=0$. In the $s$-$s_1$-$s_2$ case $n=0$ reproduces Bel-Robinson type currents \cite{Gelfond:2006be}.\footnote{In the $n=0$ case the same generating function can also be written in more conventional form as $$(\beta\gamma)^{s+1}\,z\qquad\longrightarrow\qquad C(y,\bry)C(-y,\bry)\,.$$}

\section{Fixing the 2pt Normalisations in Vasiliev's Equations Holographically}\label{normaliz}

An important subtlety to correctly interpret coefficients of currents at the equations of motion level is to determine the kinetic term normalisation for the Fronsdal fields. Such normalisation indeed play a key role to fix the cubic couplings in \cite{Sleight:2016dba}. In the following we will study these normalisations fixing the notation. Furthermore we will fix the kinetic-term normalisation for all HS fields by matching holographically the sources coming from the $0$-$0$-$s$ cubic couplings.

First of all it is important to fix the convention for spinorial indices as in \cite{Boulanger:2015ova} with
\begin{align}\label{viel}
h_\mu^{\ga\gad}h^\mu_{\gb\gbd}&= \epsilon_{\gb}{}^{\ga}\epsilon_{\gbd}{}^{\gad}\,, && h_\mu^{\ga\gad}h^\nu_{\ga\gad}=\delta^\nu_\mu\,,
\end{align}
and in Poincar\'e coordinates:
\begin{align}
h_\mu^{\ga\gad}\,dx^\mu&=\frac1{2z} \sigma_\mu^{\ga\gad} dx^\mu\,, && h^\mu_{\ga\gad}=z \sigma^\mu_{\ga\gad}\,, && g_{\mu\nu}=\frac1{2z^2} \eta_{\mu\nu}\,.
\end{align}
Here $\epsilon^{\ga\gb}=-\epsilon^{\gb\ga}$, $\epsilon^{12}=1$ and $\epsilon=i\sigma_2$ with $\sigma_i^{\ga\gad}$, $i=1,2,3$ being the Pauli matrices. Notice that in this note we work with the choice $\Lambda=2$. Fronsdal equations then read:
\begin{equation}
\frac1{4(s-1)}\,\left[\square +2(s^2-2s-2)+\ldots\right] \phi_{\ga(s)\gad(s)}=J_{\ga(s)\gad(s)}\,.
\end{equation}
where we have included the factor $\frac1{4(s-1)}$ coming from solving torsion as described in \cite{Boulanger:2015ova}. The mass term is dressed by a factor of $2$ coming from our conventions for $\Lambda$. The mapping between spinorial and vectorial indices is achieved via the vielbein as:
\begin{equation}
\phi_{\mu(s)}\equiv \phi_{\ga(s)\gad(s)}(x)\,\underbrace{h^{\ga\gad}_{\mu}\cdots h^{\ga\gad}_{\mu}}_s\,,
\end{equation}
which follows from \eqref{viel}.

The holographically reconstructed equations can be extracted from the $0$-$0$-$s$ cubic coupling:
\begin{equation}
\mathcal{L}\sim\sum_s \left[\frac{N_s^2}{2^{s+2}}\,\phi_{\mu(s)}\Box\,\phi^{\mu(s)}+\ldots\right]-\underbrace{\frac{2^4}{\sqrt{N}}}_{g}\sum_s N_sN_0^2\,\frac{2^{-\tfrac{s}2}}{\Gamma(s)}\,i^s\,\phi^{\mu(s)}\,\left(\tfrac1{2^s}\phi\,\overset{\leftrightarrow}{\nabla}_{\mu(s)}\phi+\ldots\right)\,,
\end{equation}
with an arbitrary normalisation $N_s$ for the Fronsdal kinetic term. They read:\footnote{Notice that a factor $2^{-s}$ comes from the non-canonically normalised metric in the contractions between HS field and derivatives. A further factor of $2$ comes from the non canonical normalisation for the Laplacian and another factor of $2$ comes from the variation with respect to the scalar field.}
\begin{align}
(\Box-4)\phi(x)&=g\,N_s\,\frac{2^{2-\tfrac{s}{2}}\,i^s}{\Gamma(s)}\,\phi^{\mu(s)}\,\nabla_{\mu(s)}\phi=g\,N_s\,\frac{2^{2-\tfrac{s}{2}}}{\Gamma(s)}\,\phi^{\ga(s)\gad(s)}\,C_{\ga(s)\gad(s)}\,,\label{arbitraryN}
\end{align}
where we have used that on the scalar sector of the zero form:
\begin{equation}
\tadD C=0\qquad\longrightarrow\qquad C_{\ga(s)\gad(s)}(x)=i^n\underbrace{\nabla_{\ga\gad}\cdots\nabla_{\ga\gad}}_s\,\phi(x)\,.
\end{equation}
The above equations allow to determine holographically the normalisation of the kinetic term used by Vasiliev's equations simply from the $0$-$0$-$s$ source to the scalar equation which can be extracted from the standard twisted-adjoint structure constants:
\begin{equation}
\tadD C=\omega\star C-C\star\pi(\omega)\,.
\end{equation}
Upon translating the above equations to standard Klein-Gordon equations we arrive to:
\begin{equation}
(\Box-4)\phi(x)=\frac{4}{\Gamma(s)^2}\,\phi^{\ga(s)\gad(s)}\,C_{\ga(s)\gad(s)}\,,
\end{equation}
from which comparing with \eqref{arbitraryN} we can determine the normalisation $N_s$ used by Vasiliev's equations:
\begin{align}
N_s&=\frac1{g}\,\frac{2^{\tfrac{s}{2}}}{\Gamma(s)}\,,& s&>0\,.
\end{align}
In the latter normalisation one then gets the following coupling constants for the source to the Fronsdal equations:
\begin{equation}
\left[\square +2(s^2-2s-2)+\ldots\right]\phi_{\mu(s)}(x)\,=\,\underbrace{\frac{g\,N_0^2}{N_s}\,\frac{2^{\tfrac{s}{2}+1}}{\Gamma(s)}}_{=\,2\,g^2\, N_0^2}\,i^s\left(\tfrac{1}{2^s}\phi\,\overset{\leftrightarrow}{\nabla}_{\mu(s)}\phi+\ldots\right)\,,\label{alphas}
\end{equation}
which, taking into account the normalisation \eqref{canJ} for the canonical current, determines the value for the coefficient $\alpha_s$ in \eqref{caneq1} up to a spin-independent constant:
\begin{equation}
\alpha_s=2^{1-s}g^2\,N_0^2\,.
\end{equation}
Similar results can be obtained for all other couplings of \cite{Sleight:2016dba}.
\end{appendix}

\bibliography{refs}
\bibliographystyle{JHEP}

\end{document}